\documentclass[12pt,a4paper]{article}

\usepackage{amsmath}
\usepackage{amssymb}
\usepackage{amsfonts}
\usepackage{graphicx}
\usepackage{subfig}
\usepackage{caption}
\usepackage{dcolumn}
\usepackage{bm}
\usepackage[
colorlinks,linkcolor=blue,citecolor=magenta]{hyperref}
\usepackage[top=3cm,right=2cm,bottom=2.5cm,left=2cm]{geometry}

\begin{document}
\title{Community detection based on ``clumpiness" matrix in complex networks}

\author{Ali Faqeeh\thanks{a.faqeeh@ph.iut.ac.ir}, Keivan Aghababaei Samani\thanks{samani@cc.iut.ac.ir}\\
{\it Department of Physics, Isfahan University of Technology,}\\
{\it Isfahan 84156-83111, Iran }}
\maketitle
\begin{abstract}
The ``clumpiness" matrix of a network is used to  develop a method  to identify its community structure. A ``projection space" is constructed from the eigenvectors of the clumpiness matrix and a border line is defined using some kind of angular distance in this space. The community structure of the network is identified using this borderline and/or hierarchical clustering methods. The performance of our algorithm is tested on some computer-generated and real-world networks. The accuracy of the results is checked using {\it normalized mutual information}. The effect of community size heterogeneity on the accuracy of the method is also discussed.
\end{abstract}
\tableofcontents

\section{Introduction}
Discovering the structure of complex networks and understanding the underlying laws that lead to such structures is probably one of the most important problems in complex network theory~\cite{network}. In the smallest scale (nodes) the topology of a network may be described by its degree distribution and in the largest scale global properties such as diameter of the network, mean degree, clustering coefficient,  and shortest mean path are significant. In a middle scale, properties such as community structure are important. In recent years there have been many efforts to define and find the community structure of complex networks~\cite{fortunato,self1,self2,self3}. Spectral algorithms are of most celebrated community detection methods. These algorithms usually use the eigenvalues and eigenvectors of a matrix corresponding to the network to find its community structure. The effectiveness of a method turns out to depend on the structure of the network itself. For example some methods give accurate results for networks with many loops, but are not good for tree like networks~\cite{self1}.

In this paper we use the recently introduced ``clumpiness matrix" of a network~\cite{clumpiness} to  identify its community structure. The clumpiness measure was introduced to give more information about the structure of networks and can be used as a measure to classify them. Roughly speaking the clumpiness measure shows how high degree nodes are close to each other. It turns out that the clumpiness matrix is useful in community detection of a network.

The organization of the paper is as follows. In section~\ref{s2} we introduce our method of community detection using the clumpiness matrix and give a qualitative justification for the effectiveness of the method. In sections~\ref{s3} the method is applied to some computer-generated and real-world networks with two and more than two communities. Then an improvement on the algorithm is suggested and the performance of the method is tested. Section~\ref{s4} is devoted to summary and concluding remarks.

\section{Community detection using the clumpiness matrix\label{s2}}
In this section we develop our algorithm for detecting communities using the clumpiness matrix and discuss qualitatively how the clumpiness eigenvectors can help to identify the community structure of a network.

\subsection{Clumpiness Matrix}
Clumpiness matrix($\mathbf{\Xi}$) was initially introduced by Estrada \textit{et al}~\cite{clumpiness} to extract parameters for classifying complex networks. For a network with $n$ nodes this is an $n\times n$ matrix, with  elements
\begin{equation*}
\mathbf{\Xi}_{ij}=
\left\{
\begin{array}{lcl}
k_{i}k_{j}/d_{ij}^{~2} && i\neq j \\
\\
0 && i=j\; ,
\end{array}
\right.
\end{equation*} where $k_{i}$ is the degree of node $i$ and $d_{ij}$ is the magnitude of the shortest path between nodes $i$ and $j$. Let $\mathbf{V_{1}},\mathbf{V_{2}}, ..., \mathbf{V_{n}}$ denote the eigenvectors corresponding to the (non-increasing) sequence of eigenvalues $\varepsilon_{1},\varepsilon_{2}, ..., \varepsilon_{n}$ of the clumpiness matrix respectively. Since the elements of $\mathbf{\Xi}$ are non-negative; elements of $\mathbf{V_{1}}$ are all of the same sign according to the Perron-Frobenius theorem. Therefore, the elements of $\mathbf{V_{2}}$ can be positive or negative in order to satisfy orthogonality condition. Element $v_{i}(q)$ of $\mathbf{V_{i}}=\left( v_{i}(1),v_{i}(2),...,v_{i}(q),...,v_{i}(n)\right)$ corresponds to the node $q$ of the network. In networks with two communities, observations show that nodes with positive elements of $\mathbf{V_{2}}$ belong to the same community. This is also the case for negative elements. However, the border between these two communities is not exactly zero. In other words in some cases a node with a negative element belongs to the community of nodes with positive elements or vice versa. Hence, in order to detect communities, we need a more strict method rather than merely considering the sign of elements of $\mathbf{V_{i}}$s. This, and its generalization to networks with more than two communities is explained in next subsections.

\subsection{Networks with two communities}
\begin{figure} \centering
\subfloat[]
{\includegraphics[width=70mm]{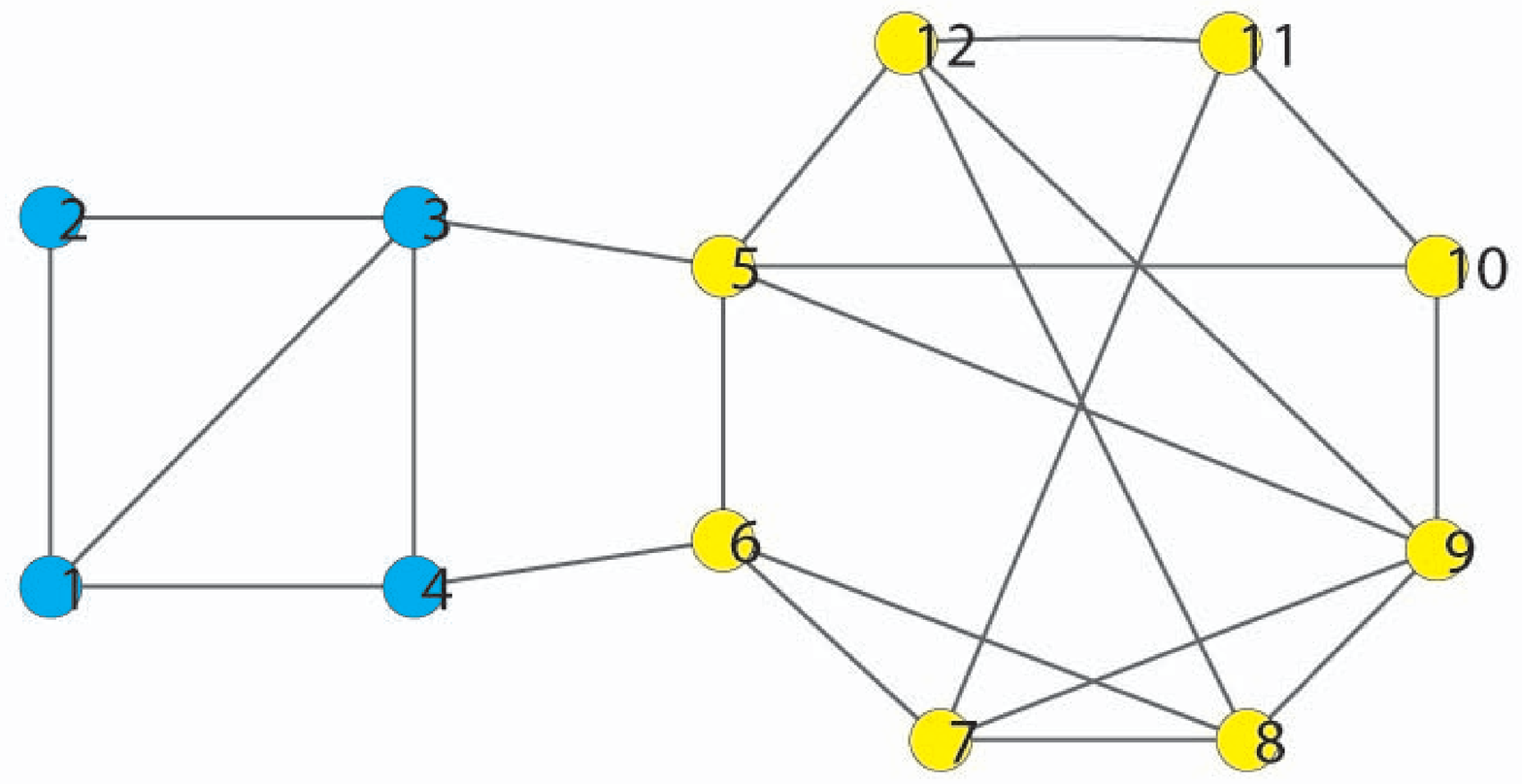}\label{fsa}}
\qquad
\subfloat []
{\includegraphics[width=50mm]{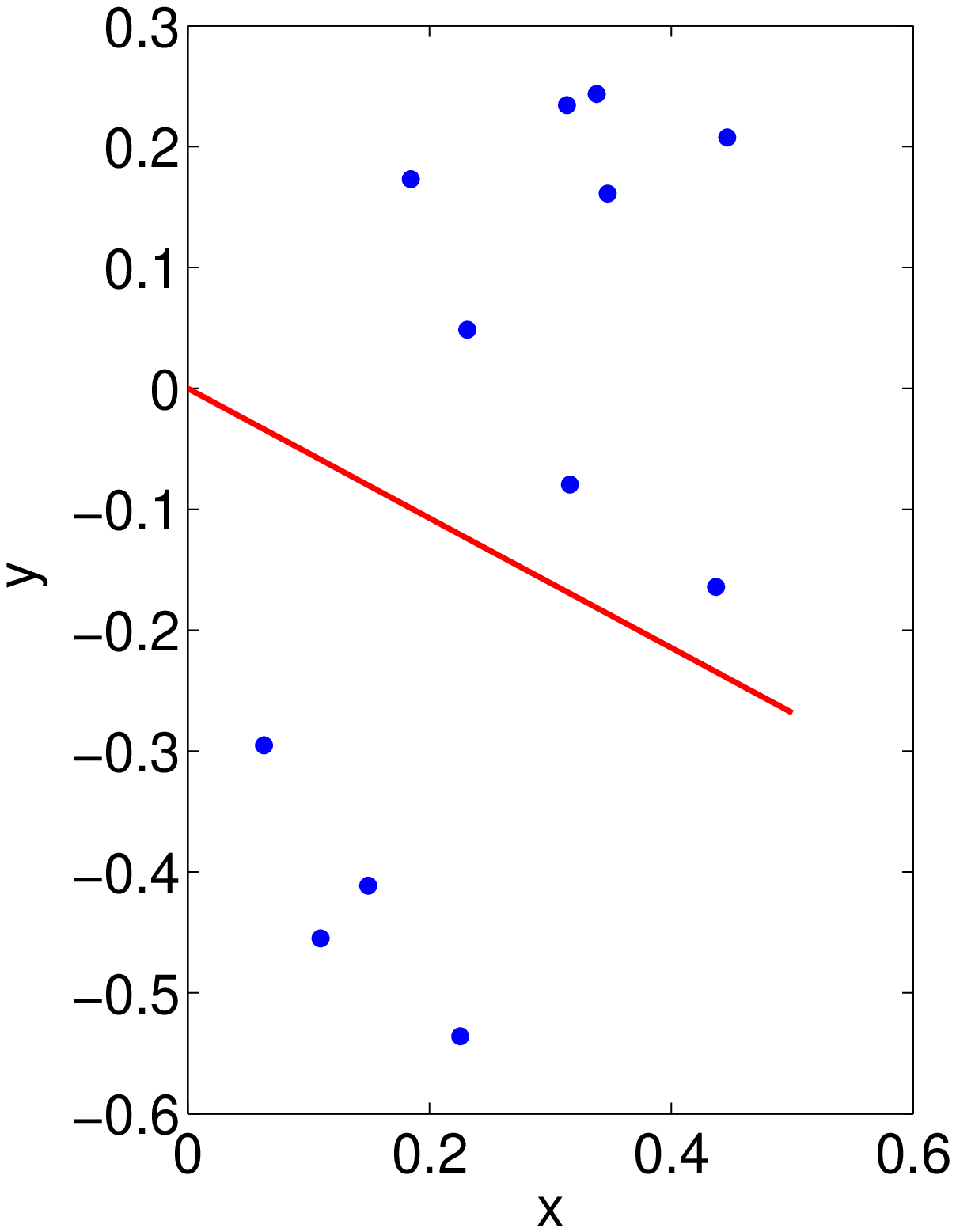}\label{fsb}}
\caption{(a) A graph with two communities. (b) Projection space of the graph. The representative points and the correct borderline is plotted.}\label{fs}
\end{figure}

For a network with two communities, we deem a two dimensional space (the so called {\it projection space}). The node $i$ of the network is shown in this space by $(x,y)=(v_{1}(i),v_{2}(i))$. A similar projection space was previously employed by Donetti and Mu\~{n}oz using the Laplacian matrix to unfold the community structure of complex networks \cite{Donetti} \footnote{For more information about the spectral clustering and properties of projection space interested reader can study Ref.~\cite{SC}.}. We use the same notion for the clumpiness matrix. Projecting all nodes of the network in this space, it is observed that usually they are accumulated in two branches (see FIG.~\ref{fs}). Each group of points in this space is supposed to indicate one community. Therefore, one should show a way to divide the points into two groups. In this plane each node is alternatively identified by its polar coordinate $(r,\theta)$, where $r=\sqrt{x^{2}+y^{2}}$ and $\theta=\arctan(y/x)$. Our observations show that the component $\theta$ of each node is appropriate to identify its community. In other words, the $\theta$ components of nodes of each community are  close to each other. Hence, we need a borderline with angle $\Theta$, to separate the two branches. We take into account four possibilities to define $\Theta$ :
\begin{equation}\label{AA}
    \Theta_{AA}=\frac{1}{N}\sum\theta_{i}\; ,
\end{equation}
\begin{equation}\label{MA}
     \Theta_{MA}=\frac{1}{2}(\theta_{i}^{max}+\theta_{i}^{min})\; ,
\end{equation}
\begin{equation}\label{MH}
     \Theta_{MH}=\frac{1}{2}(\theta_{y_{i}^{max}}+\theta_{y_{i}^{min}})\; ,
\end{equation}
\begin{equation}\label{WA}
    \Theta_{WA}=\frac{\sum|y_{i}|\theta_{i}}{\sum|y_{i}|}\; ,
\end{equation} where $ \theta_{i}^{max}$ and  $\theta_{i}^{min}$ are maximum and minimum angles of points respectively, and $\theta_{y_{i}^{max}}$ and $\theta_{y_{i}^{min}}$ are angles of points with maximum and minimum ordinates respectively.
The quantity $\Theta_{AA}$ is the average angle and $\Theta_{MA}$ is the maximum-minimum average. Observations show that using some combination of angles and ordinates of points to define the borderline leads to better results in community detection. This is the motivation to define borderlines $\Theta_{MH}$ and $\Theta_{WA}$. Eq.~(\ref{WA}) is a weighted averaging on angles of points, where weights are their ordinates, and it turns out that this gives better results than other definitions. An example of a simple graph having two communities is shown in FIG.~\ref{fs}.

\subsection{Networks with more communities}
In networks with more than two communities, the use of eigenvectors' relation is somehow different. The angle associated to each point $i$, which is $\arctan(\frac{\mathbf{V_{2}}(i)}{\mathbf{V_{1}}(i)})$, separates one community from another when there are two of them to be partitioned. This inspires us to assume that the ratio of next eigenvectors to $\mathbf{V_{1}}$ (eigenvector with greatest eigenvalue), can be used to find more communities. Suppose that the network is constructed from $C$ communities, we use the first $C$ eigenvectors of the clumpiness matrix to construct an $n\times (C-1)$ matrix, $\mathbf{\Gamma}$, in which the column $j-1$ includes angles obtained from $\arctan(\frac{\mathbf{V_{j}}(i)}{\mathbf{V_{1}}(i)})$. That is:
\begin{equation}
\mathbf{\Gamma}_{i,j-1}=\arctan\left(\frac{\mathbf{V_{j}}(i)}{\mathbf{V_{1}}(i)}\right),~~~~j =2,..., C.
\end{equation}
Depicting points in a $C-1$ dimensional projection space constructed from $C-1$ columns of $\mathbf{\Gamma}$, the points related to each community are accumulated in distinct clusters. This aggregation of depicted points has also a physical justification which is explained in subsection~\ref{QA}. In fact the angle $\mathbf{\Gamma}_{i,j-1}$ gives the direction of point $i$ with respect to a reference direction (i.e. direction of $\mathbf{V_{1}}(i)$) in each $\mathbf{V_{1}}-\mathbf{V_{j}}$ plane. Now a better result could be achieved if we calculate the angles between every two points without using such a reference and find accumulation of nodes with the same direction. In this case, we should use a C-dimensional space constructed of rows of matrix $\mathbf{U} $:
\begin{equation}
\mathbf{U}_{i,j}=\mathbf{V_{j}}(i) ,~~~~j =1,..., C.
\end{equation}

In order to partition the resulted clusters we used agglomerative hierarchical clustering algorithm \cite{Wasserman,network}. This algorithm starts with each point as a separate cluster and then progressively merge the points into bigger and bigger groups.  Different methods such as simple-linkage, complete-linkage and average-linkage could be used to determine when two groups should be merged together. Here we used the average-linkage method as it led to better results. The rest of the procedure depends on choosing $\mathbf{\Gamma}$ or $\mathbf{U}$.

In the first case the Euclidean distance between points in the projection space should be derived from matrix $\mathbf{\Gamma}$. Using the yielded distances, we build a cluster tree according to points' average proximity. Then, the $C$ clusters associated with communities can be derived. In case of considering the other matrix for projection space, we are supposed to treat points as vectors and use angular distances instead.

Note that it is feasible to apply this method to the case of networks with two communities too. However our observations show that defining the borderlines leads to a better performance in that case. On the other hand, in the higher dimensions of projection space it is not possible to define such a borderline or generalize it without a considerable loss in performance and accuracy of the method. Moreover, the use of this method in two dimensions helps to understand how the method works.

It is worth mentioning that we need the number of communities to proceed with our algorithm. In some cases, a priori knowledge about this number exists. Otherwise we should obtain it somehow. A possible way (inspired by Donetti and Mu\~{n}oz \cite{Donetti} and also used in \cite{self1}) is to calculate modularity \cite{Newman} while enlarging the dimension of the projection space i.e. using two and then more eigenvectors. One can also use the gap between eigenvalues of directed Laplacian matrix \cite{self1}. This can be done rather fast by Lanczos algorithm \cite{Lanczos} since a finite number of eigenvalues should be obtained. In addition, there are algorithms devised for this common problem \cite{Still,Tibshirani}. In our tests we do not take account of this problem and use the given number of communities.

\subsection{Qualitative analysis\label{QA} }
The outcome of this method can be explained by looking into eigenvectors and their mutual relations. Let speculate the network is consisted of nodes that interact with each other through hopping of some entities. Then one can correspond the following Hamiltonian to the network:
\begin{equation}\label{Hamiltonian}
H=\sum_{i,j}g_{ij}(|i\rangle\langle j|+|j\rangle\langle i|)+\sum_{i}W_{i}(|i\rangle\langle i|)\; ,
\end{equation}
where $W_{i}$ acts like a localized potential which is  considered to be zero in this paper for simplicity. It is rational to consider $g_{ij}$ proportional to $-\frac{k_{i}k_{j}}{d_{ij}^{~2}}$. That is, hopping between sites $i$ and $j$ is proportional to number of exiting ways from node $i$ multiplied by number of entries to node $j$, and inversely proportional to square of nodes' distance.  Hence, clumpiness matrix can be viewed as the negative of the Hamiltonian of the system. This Hamiltonian describes a system in which each site (corresponding to a node of the network) interacts with all other sites. Therefore, $V_1$ would be the ground state of the system, which is also the most ordered state, i.e. all components of $V_1$ are positive. By increasing the energy of the system its initial order diminishes. The first exited state, $V_{2}$, encompasses two regions of opposite spin-like states. The regions are expected to coincide with communities just in case of a network with two communities, since there are more and stronger interactions among the members in each group than interactions between nodes in different groups. For networks with more than two communities one should take into account more eigenvectors (i.e. more excited states).  When the system gradually advances to higher states, interactions make  sites in each community to behave alike  each other. In the $C$th level, short range order is held in $C$ distinct regions.
\begin{figure} \centering
\includegraphics[width=80mm]{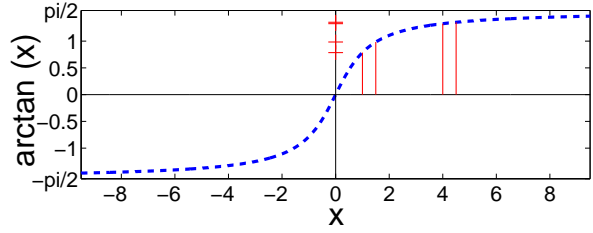}
\caption{\label{fatan}Obtaining the inverse tangent of some data is like mapping them from one infinite space to a finite one. }
\end{figure}

These regions are more distinguishable if we compare their order with the order of the ground state. This is best achieved taking into account the ratio of eigenvector components of each level to those of the ground state $(V_{j}/V_{1})$. By taking the inverse tangent of these ratios, we restrict the maximum values, making them closer together (FIG.~\ref{fatan}). Accordingly, columns of matrix $\mathbf{\Gamma}$ tell us how sites have changed with respect to others in each level. As a result, clusters of points in $C-1$ dimensional space indicate the most related nodes, defining the optimum borders of regions.

\begin{figure} \centering
\scalebox{.5}{\includegraphics {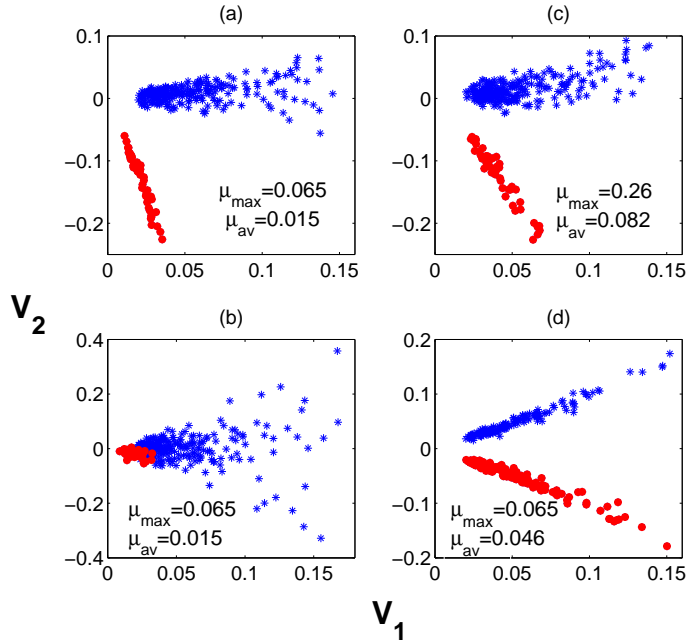}}
\caption{\label{sres} Samples of different situations on several realizations of a benchmark network with two communities. In (a)-(c) the community sizes are 50 and 250, and in (d) both communities have 150 nodes.}
\end{figure}

For networks with heterogeneous community size distribution the situation may differ. In a very small community, nodes have lower degrees (i.e. fewer connections) with respect to nodes in a large community. For small mixing between communities, nodes in the small community are more connected to the peer nodes with lower degrees, while nodes in the big community have many connections with high degree nodes. Hence, nodes in small communities have low valued components in  corresponding eigenvectors, i.e. they are less occupied, with respect to others in large communities which could have a broad range of component values in corresponding eigenvectors. As a result, the branch related to a small community will become much shorter and its points accumulate near the origin so that they all mix with the points related to the large communities. On the other hand, for large mixing between communities, more connections between nodes of different communities will adjust the heterogeneity and the size effect will be reduced. FIG.~\ref{sres} shows examples of this situation. 

The networks in FIG.~\ref{sres} are constructed using an algorithm inspired from LFR benchmark~\cite{benchmark}. This algorithm generates networks with power-law degree distribution and two communities in which mixing parameter and the sizes of communities are arbitrary. The maximum and average mixing parameter ($\mu$) is indicated in each case. FIG.~\ref{sres}a and FIG.~\ref{sres}b are samples of the projection space ($\textbf{U}$) for two different situations happened in a number of realizations of a network having communities with $50$ and $250$ nodes and low mixing parameter between them. The degree distribution power is $-2$ and minimum and maximum degrees are $20$ and $100$ respectively. In FIG.~\ref{sres}a the points are well separated in two distinct branches. On the other hand, in FIG.~\ref{sres}b the bigger branch is extended while points of the small community are accumulated near the origin. In $50$ realizations of the network $8$ of them was related to this situation. For a larger mixing parameter (FIG.~\ref{sres}c) this situation is less probable ($1$ out of $50$). The two communities in FIG.~\ref{sres}d have the same size. This, results in perfect formation of two branches even for small mixing parameter in all $50$ realizations.

One can also use a similar argument about the performance of Laplacian matrix on detecting the communities. In this regard, there is a difference between behaviours of Laplacian and clumpiness matrices. This difference results from the type of interaction that each matrix represents. The clumpiness matrix assumes a global interaction between the nodes, on the other hand the Laplacian matrix represents interactions between the neighbors plus a potential well in each site (i.e. the diagonal elements). Because of the tight-binding interactions in the Laplacian matrix, the occupation of each network's site (i.e. element of an eigenvector) only depends on its adjacent sites. In small mixing parameters this leads to good separation of communities. However, when the number of outer and intra links are comparable, the elements of eigenvectors will be very close and mixed together so that dividing the clusters is hard.

On the other hand, due to the infinite range of interactions in clumpiness matrix, each site is occupied according to its interaction with all the network. The nodes with more connections and stronger interactions are more occupied. Hence, these nodes are placed at the end of the branches corresponding to their communities in the projection space. By increasing the mixing parameter the branches approach each other; this approach is more appreciable near the origin than at the end of the branches. Therefore, in larger mixing parameters the nodes which are far from the origin can be well separated. Additionally, the nodes which are not close to boundaries of two branches are still separable. This is in contrast with the Laplacian case in which the difference between different sites is not appreciable. This results in better performance of clumpiness matrix in large mixing parameters.

\subsection{Time complexity of the method}
The running time of our algorithm depends on the size of the system. Below we briefly discuss this dependence for different parts of the algorithm. 

In order to obtain the clumpiness matrix one needs to calculate all shortest paths of the network. Two popular algorithms for this purpose are Floyd-Warshall with $\mathcal{O}(n^3)$ time complexity and Johnsons with $\mathcal{O}(n^2log(m))$ complexity, with $n$ and $m$ being the number of nodes and links respectively \cite{Cormen,Sedgewick}. The average-linkage hierarchical clustering method which is used in our algorithm has a time complexity of $\mathcal{O}(n^2log(n))$~\cite{HC}. The complete-linkage method used by Donetti \textit{et al}~\cite{Donetti} has the same complexity, but according to our observations average-linkage leads to better results. Finally we need to obtain a finite number of clumpiness matrices' eigenvectors corresponding to its largest eigenvalues. This can be done by algorithms based on Lanczos method ~\cite{Lanczos} which their running time have a $\mathcal{O}(n^2)$ dependency on the size of the matrix~\cite{Amati}. The dependency of this method on the number of iterations and eigenvectors is negligible as they are finite. Hence, the time complexity of our algorithm is $\mathcal{O}(n^2 log(m))$.

\subsubsection*{Normalized Mutual Information}
Normalized mutual information ($I(A,B)$)\cite{Information}, which provides a good inspection on the discrimination power of detection methods, is based on intersection matrix $\mathbf{N}$. The entry $\mathbf{N}_{ij}$ of $\mathbf{N}$,  equals to the number of nodes in detected community $j$ common to real community $i$. In a network with the number of built in communities $c_{A}$, let $c_{B}$ be the number of  detected communities. Then $I(A,B)$ is defined as below:
\begin{equation}\label{Normalized Mutual Information}
I(A,B)=\frac{-2\sum_{i=1}^{c_{A}}\sum_{j=1}^{c_{B}}N_{ij}\ln(N_{ij}N/N_{i.}N_{.j})}{\sum_{i=1}^{c_{A}}N_{i.}\ln(N_{i.}/N)+\sum_{j=1}^{c_{B}}N_{.j}\ln(N_{.j}/N)}\; ,
\end{equation} where $N$ is the total number of nodes, $N_{i.}$ the sum over the row $i$ and $N_{.j}$ the sum over column $j$ of $\mathbf{N}$. In our method we suppose that the number of communities are given initially, therefore $c_{A}$ and $c_{B}$ are equal. The normalized mutual information is a number between zero and one, and greater values of $I(A,B)$ mean the better community detection.

\section{Computational results\label{s3}}
In this section we apply our community detection method on some artificial and real-world networks.
\subsection{Networks with two communities}
\subsubsection{Artificial networks}
\begin{figure} \centering
\subfloat[Network]
{\includegraphics[height=60mm]{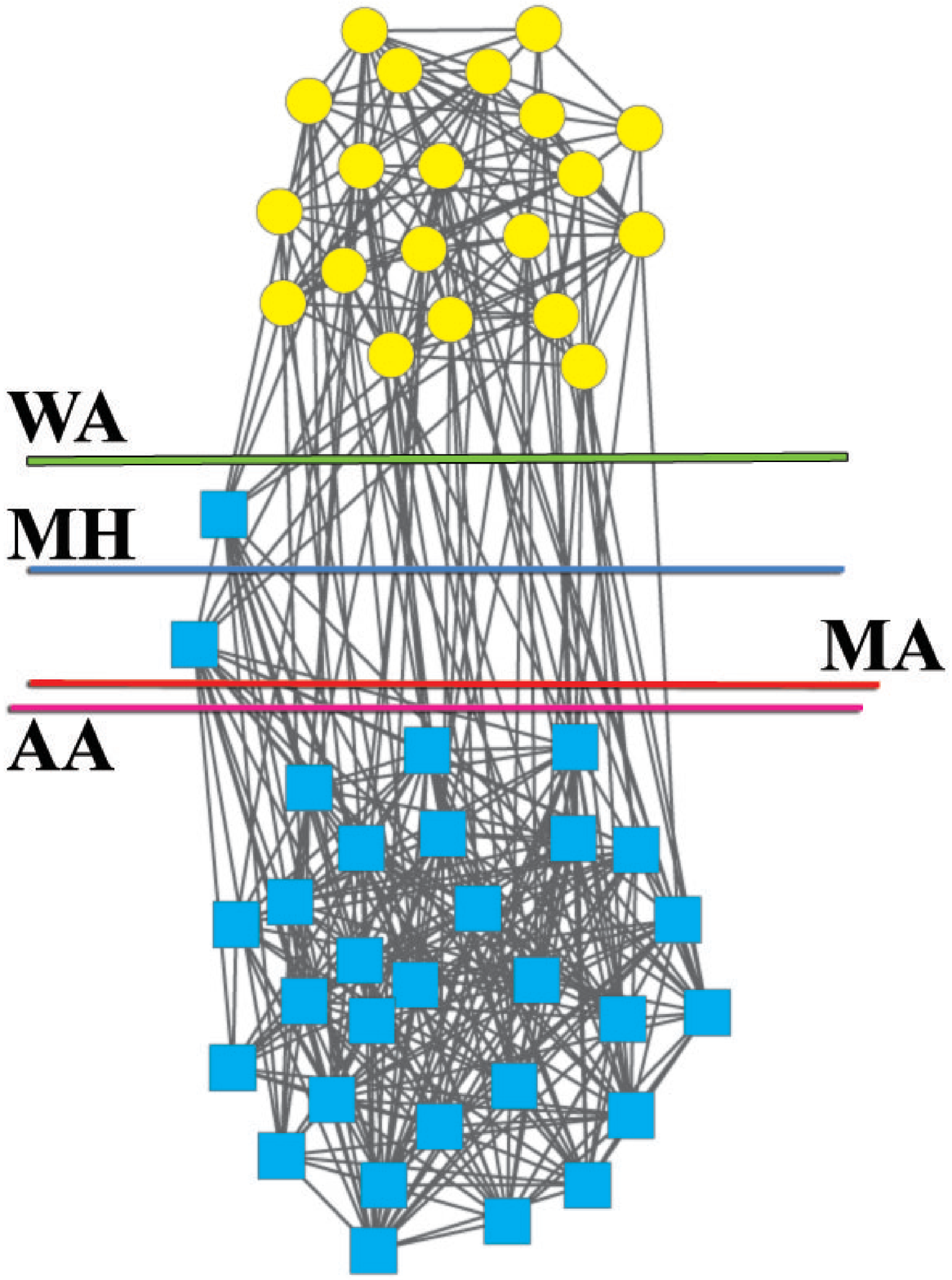}\label{fra}}
\qquad
\subfloat [Points accumulation]
{\includegraphics[height=60mm]{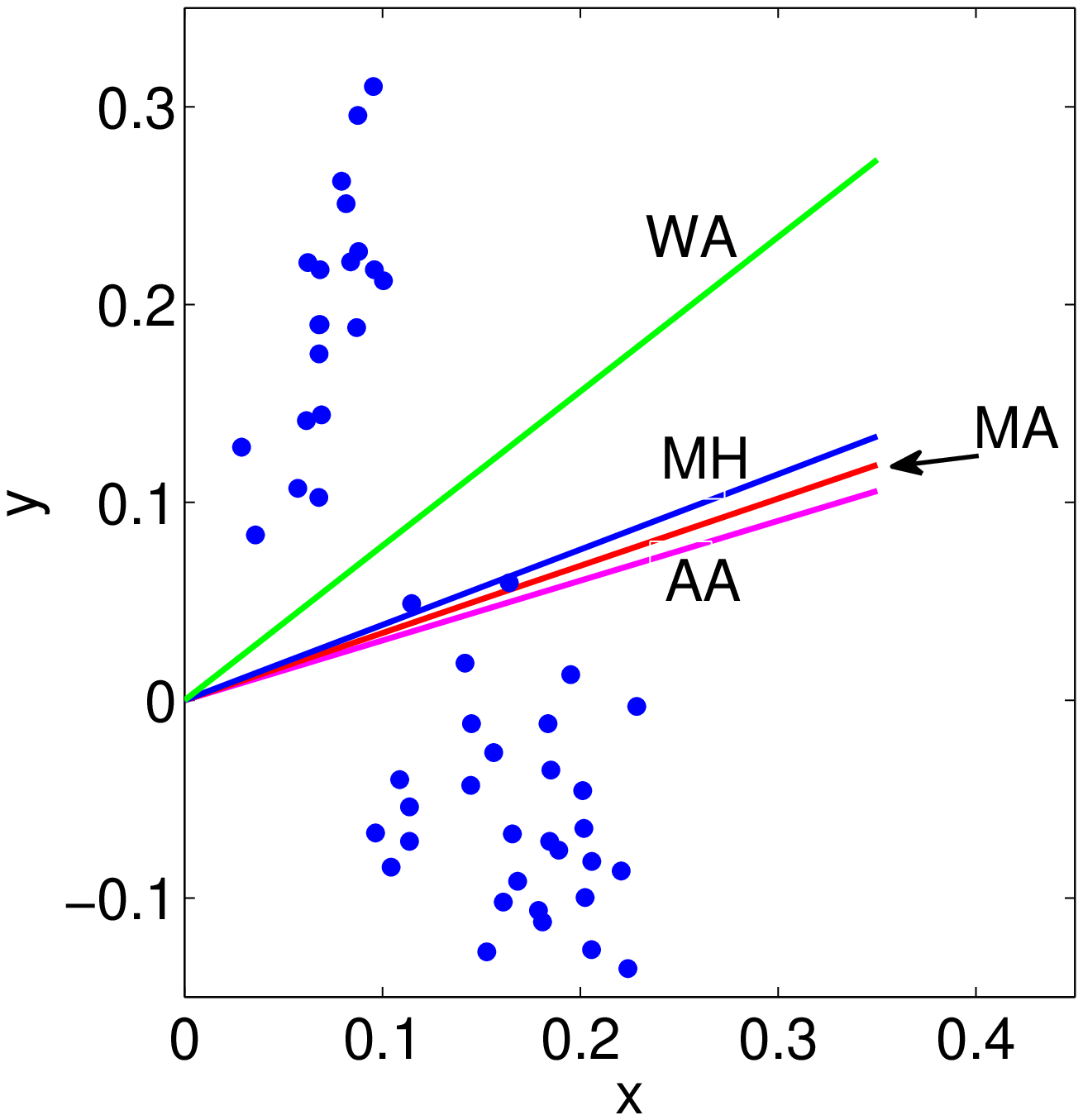}\label{frb}}
\caption{(a) An artificial graph with two communities. (b) Projection space }\label{fr}
\end{figure}

Our first example (FIG.~\ref{fra}) is a network with two communities having 30 and 20 nodes respectively. The connection probability between the members of each community is 0.5 and for those in different group is 0.1. FIG.~\ref{frb} shows the projection space of this network with the borderlines. FIG.~\ref{frb} illustrates that the borderline WA identifies the communities more accurately. Borderlines MH and MA have one misidentification, while AA has 2 misidentification.
\begin{figure} \centering
\scalebox{.45}{\includegraphics{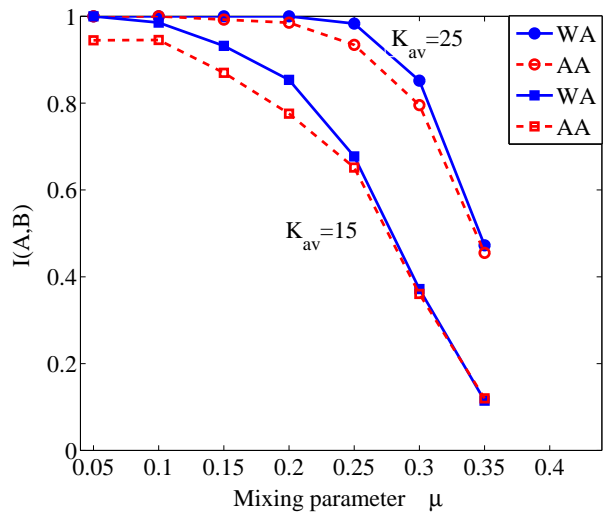}}
\caption{\label{fm2c}Variation of normalized mutual information against mixing parameter for networks with 2 communities with average degrees $25$ and $15$. The other network parameters are as follows: $N=100,~\gamma=3,~K_{max}=50,~S_{min}=30$, and $S_{max}=60$.}
\end{figure}

By applying our method on benchmark networks~\cite{benchmark}, its accuracy  will be observed more precisely. On benchmark networks, we can analyze the boundary of accuracy by calculating the variation of normalized mutual information with respect to different network parameters. In FIG.~\ref{fm2c} the variation of normalized mutual information with mixing parameter $\mu$ (the ratio of inter to intra links) for networks with two communities is depicted. The benchmark networks have $100$ nodes and the initial parameters are defined as follows: degree distribution exponent $\gamma=3$, maximum degree $K_{max}=50$, minimum community size $S_{min}=30$ and maximum community size $S_{max}=60$. In this figure we can see the average $I(A,B)$ for WA and AA definitions. The two other definitions have $I(A,B)$ between WA and AA lines. We see that the discrimination power decreases with the increase of $\mu$. It should be noted that in these benchmark networks for $\mu\gtrsim0.4$ the communities are not well defined any more, according to the planted $\ell$-partition model~\cite{l-partition,CDA}.

\subsubsection{Real networks}
\begin{figure} \centering
\subfloat[]
{\includegraphics[height=60mm]{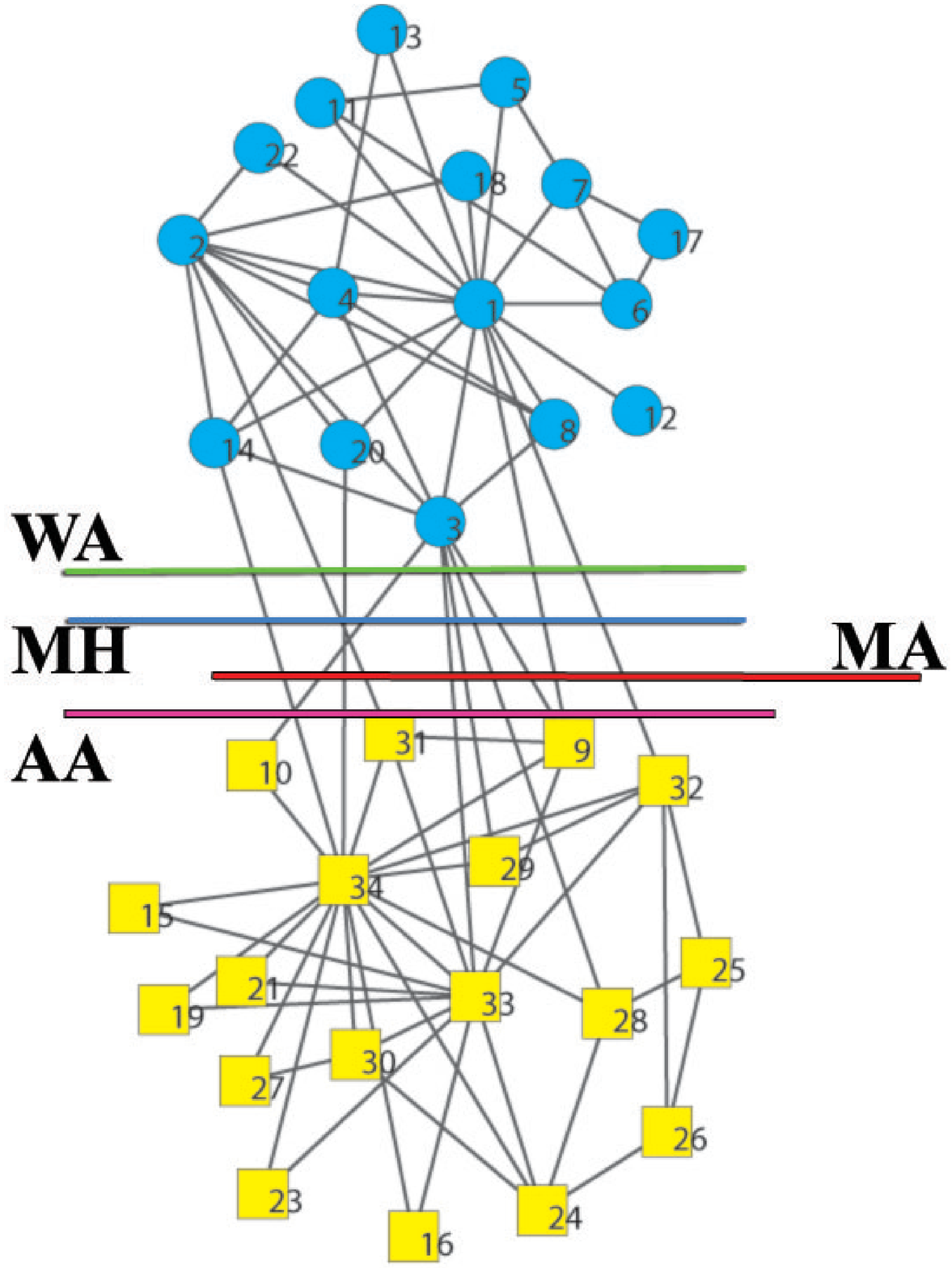}\label{fka}}
\qquad
\subfloat []
{\includegraphics[height=60mm]{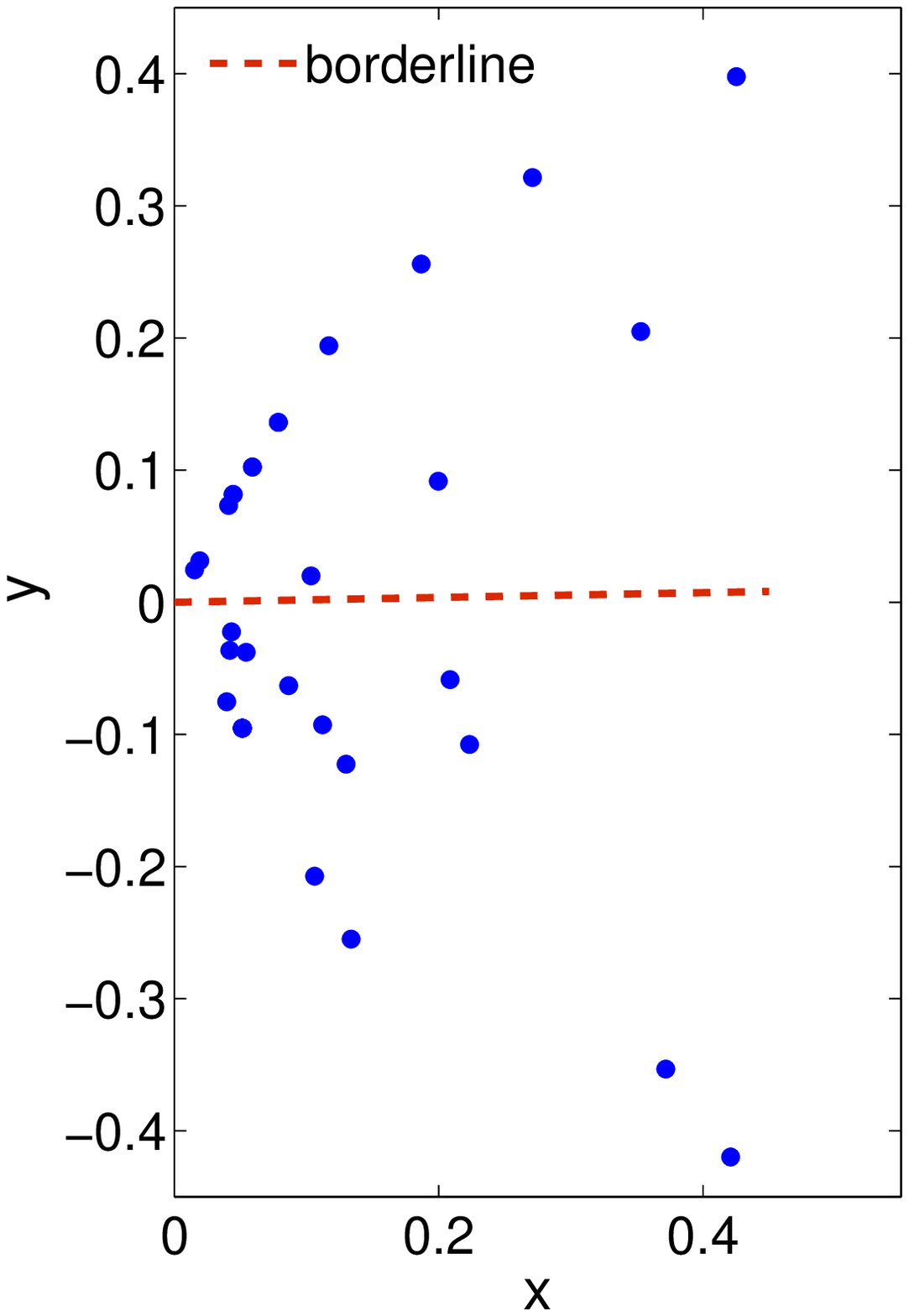}\label{fkb}}
\caption{(a) Karate Club network. (b) Projection space. All four definitions for $\Theta$ yield the same borderline and detect the two communities correctly}\label{fk}
\end{figure}

\begin{figure} \centering
\subfloat[]
{\includegraphics[height=60mm]{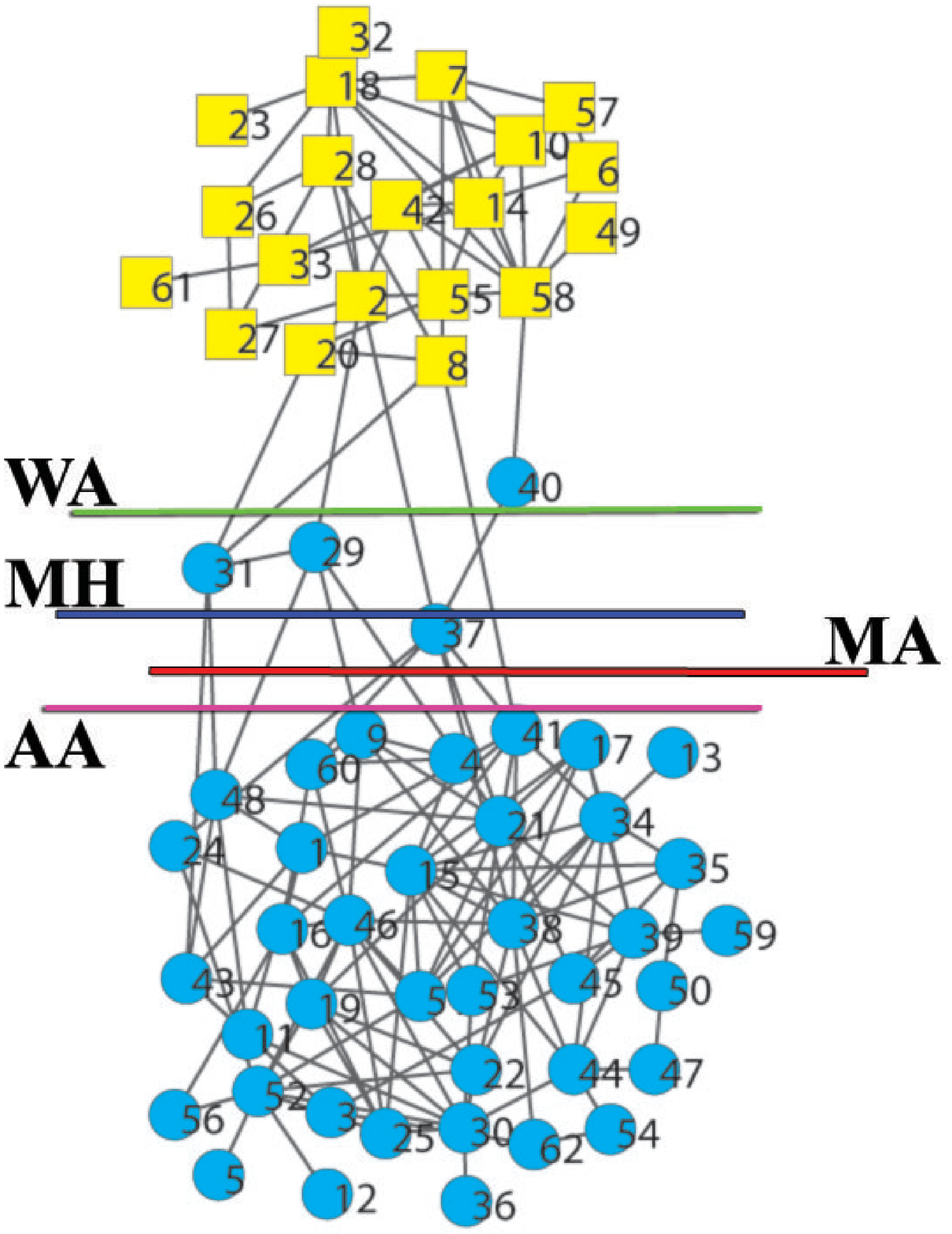}\label{fda}}
\qquad
\subfloat[]
{\includegraphics[height=55mm]{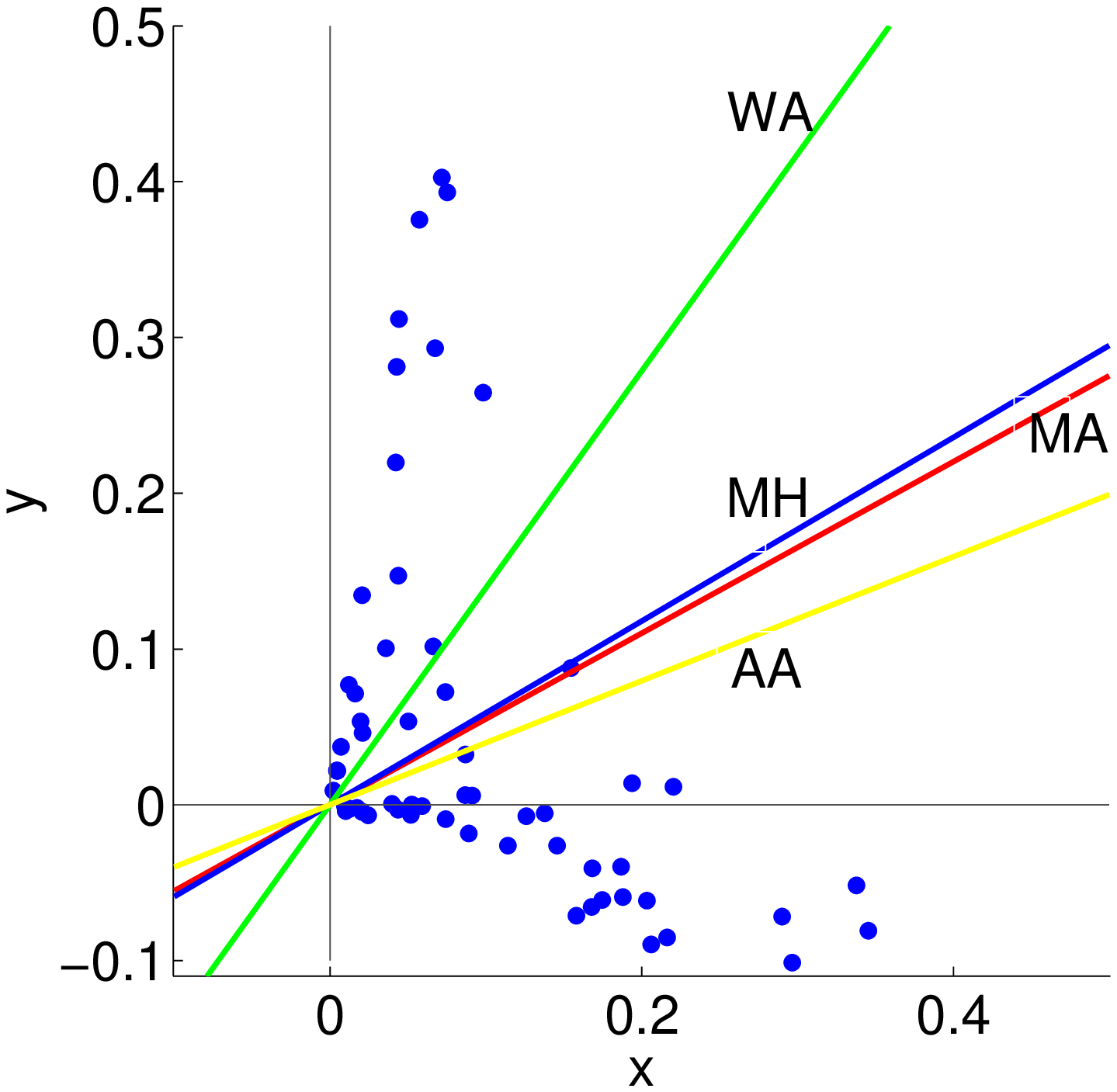}\label{fdb}}
\caption{(a) Dolphin social network. (b) Projection space. WA has the best result with only one misidentification\label{fd}}
\end{figure}

In addition we applied our method to the real networks of Zachary's karate club (FIG.~\ref{fk})~\cite{karate} and dolphin social network (FIG.~\ref{fd})~\cite{dolphin}. There is no preference among the four borderlines in karate club network. However, in dolphin social network, the weighted average (WA) seems to work more properly, while the usual average (AA) has the lowest distinction power.

\subsection{Networks with more communities}
\subsubsection{Test on the benchmark networks}
\begin{figure} \centering
\subfloat[]
{\includegraphics[height=60mm]{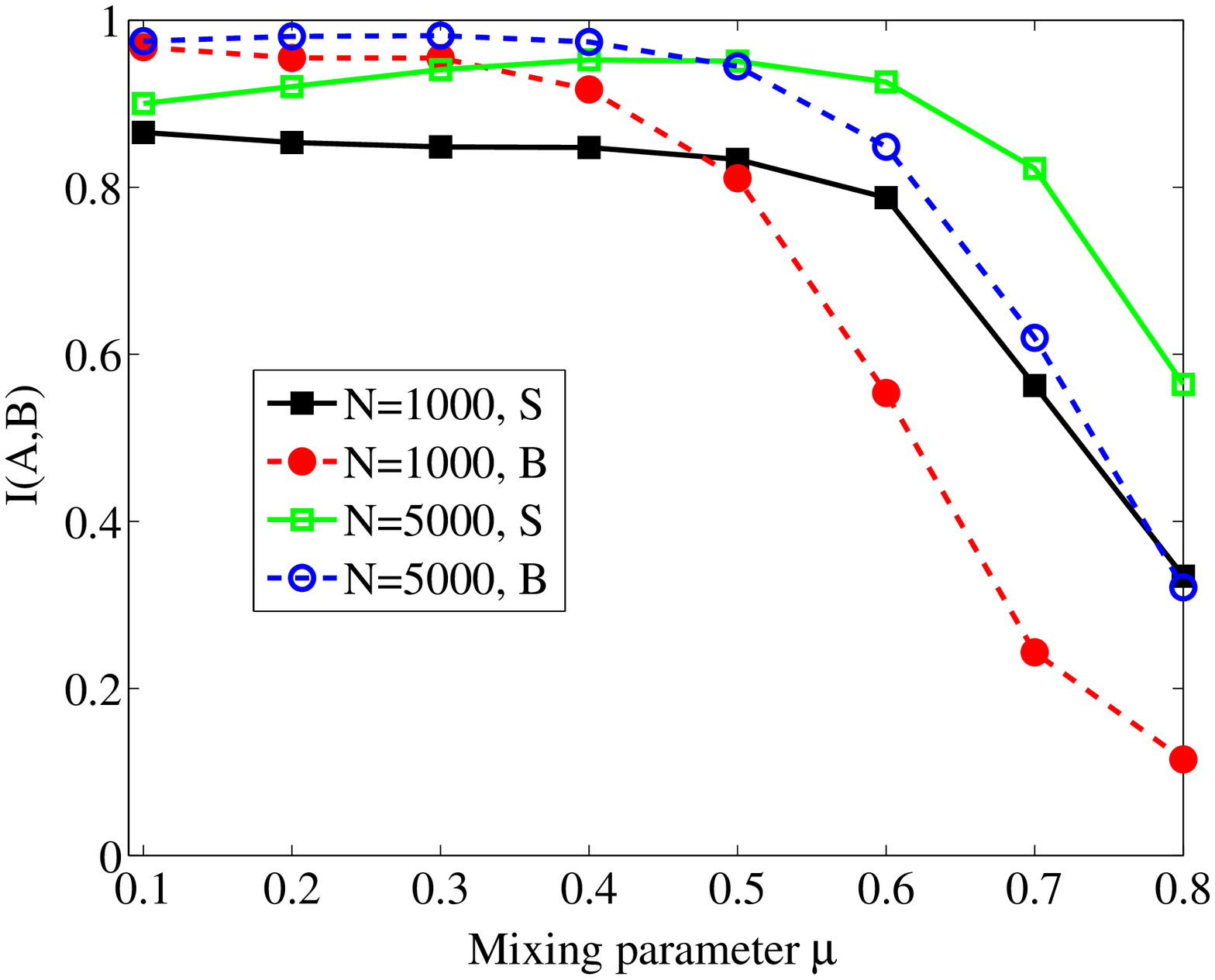}\label{fba}}
\qquad
\subfloat[]
{\includegraphics[height=60mm]{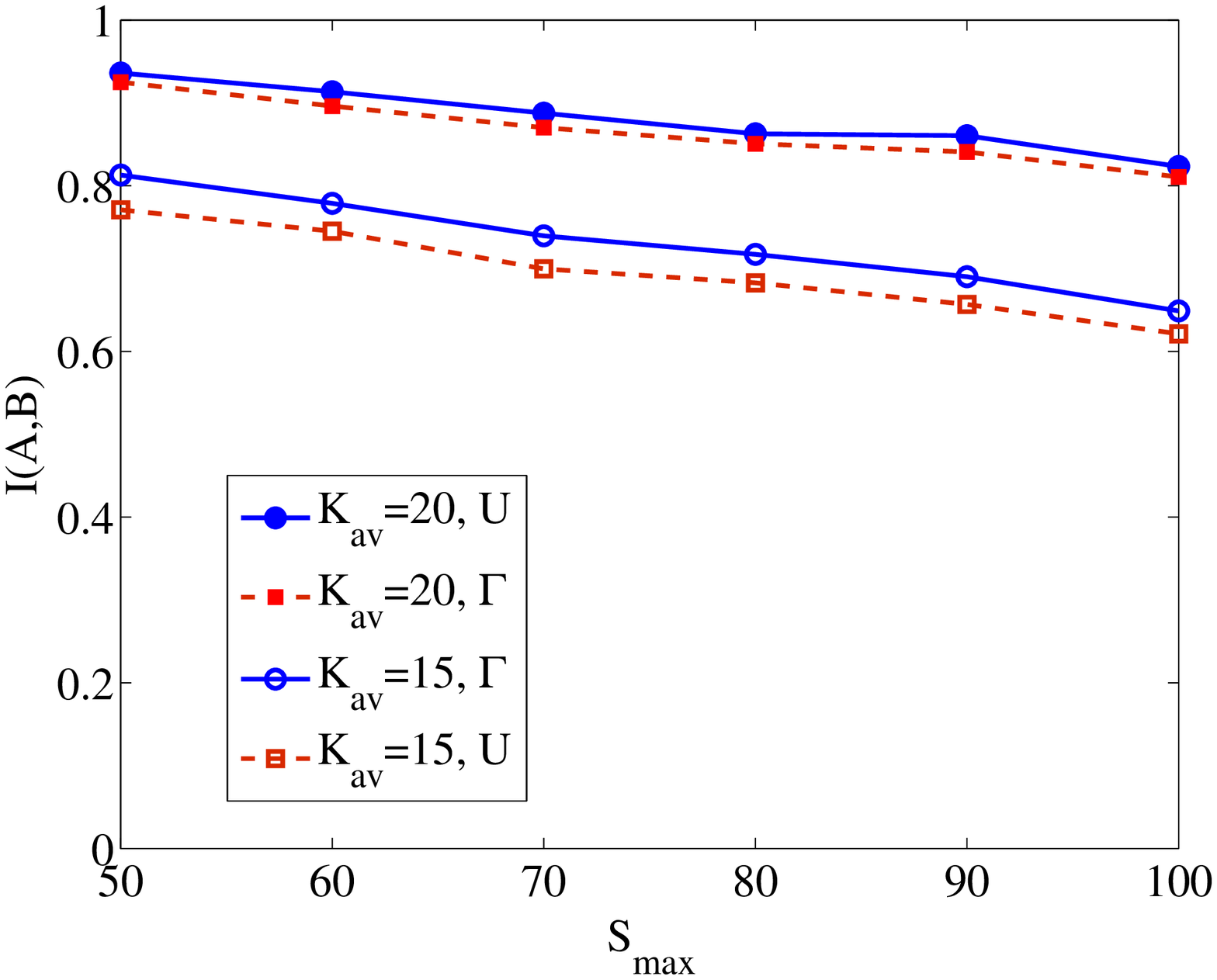}\label{fbb}}
\caption{(a) Variation of normalized mutual information against mixing parameter in networks with $1000$ and $5000$ nodes, resulted from using \textbf(U) as projection space. (b) The effect of community size heterogeneity on $I(A,B)$ in networks with 1000 nodes. The other network parameters are: $\gamma=2,~\beta=1,~K_{max}=50$, and in (a) \textbf{S} stands for $(S_{min}=10,~S_{max}=50)$ and \textbf{B} for $(S_{min}=20,~S_{max}=100)$.}\label{fb}
\end{figure}

We used LFR benchmarks \cite{benchmark} to examine the performance of the algorithm on networks with different parameters. FIG.~\ref{fba} illustrates the performance of algorithm on networks with the average degree $K_{ave}=20$, the power-law exponent of degree distribution exponent $\gamma=2$ and community size distribution $\beta=1$. Networks with $1000$ and $5000$ nodes are considered. In addition, two ranges are considered for minimum and maximum community sizes: averagely small community sizes $(S_{min}=10,~S_{max}=50)$ and averagely big community sizes $(S_{min}=20,~S_{max}=100)$ which are indicated by signs $S$ and $B$ respectively.

A better performance is observed for larger networks ($N=5000$) according to normalized mutual information $I(A,B)$. In case of smaller communities and larger networks $\lbrace N=5000,S\rbrace$, $I(A,B)$ slightly increases with $\mu$ up to $\mu\leq0.5$,. This is because of the fact that size resolution effect (resulted from large number of very small communities with average degree fewer than $K_{ave}=20$) decreases when mixing parameter grows. For $\mu\geq0.6$, the high mixing between communities dominantly makes it harder to detect communities, causes the decrease in $I(A,B)$. 

In FIG.~\ref{fbb}, the size resolution effect is investigated for networks with $1000$ nodes and different average degrees ($\mu=0.5$, $\gamma=2$, $\beta=1$) by considering $S_{min}=20$ and plotting $I(A,B)$ against $S_{max}$. The results are depicted for different choices of projection space ($\textbf{U}$ or $\mathbf{\Gamma}$). 

\subsection{Improving the method}
The hierarchical clustering algorithm constructs the communities according to the proximity of points in the projection space. In each step of this algorithm the two closest clusters of points are merged together, where the closeness of each pair of clusters is defined according to a similarity measure
\footnote{In the group-average clustering method we used, the similarity of two clusters equals the average of distances between all pairs of points in the two clusters.}.
We are interested in a merging procedure in which there is a priority for joining clusters that are also closer to each other in the real network. In their spectral algorithm, Donetti and Mu\~{n}oz \cite{Donetti} take this into account by inserting a constraint that two clusters are merged only if there is a link between them in the network. Implementing this constraint will improve the performance of our algorithm as well. However our observations show that a considerably better performance is achieved by simply altering the initial similarity between pair of points as follows:
\begin{equation}
Sim(i,j)=Sim(i,j)*d_{i,j}^{~2}.
\end{equation}
Here $Sim(i,j)$ is the similarity between points $i$ and $j$ and $d_{i,j}$ the length of the shortest path between nodes $i$ and $j$. This change does not affect the main body and speed of the algorithm.

\subsubsection{Performance of the method on LFR benchmarks\label{ssp}}
\begin{figure} \centering
{\includegraphics[height=80mm]{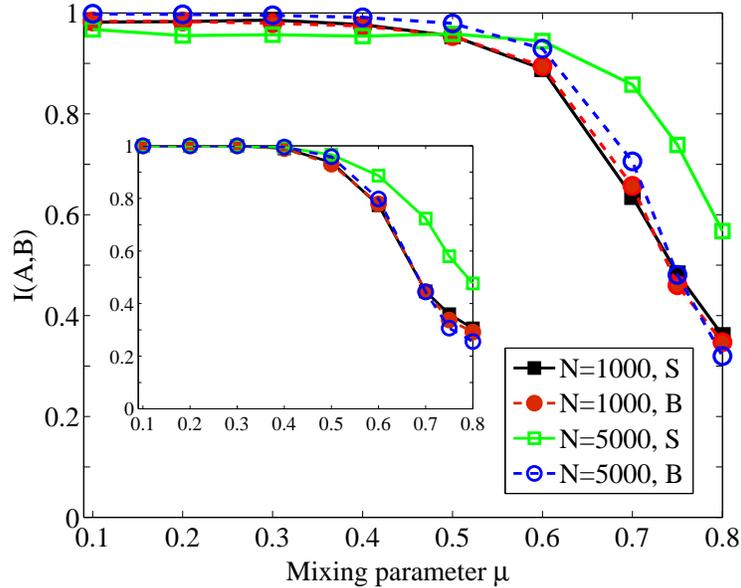}}
\caption{Performance of our algorithm against mixing parameter in networks with $1000$ and $5000$ nodes by using clumpiness matrix (main figure) and using Laplacian matrix (inset). The other network parameters are as follows: $\gamma=2,~\beta=1,~K_{max}=50$ and \textbf{S} stands for $(S_{min}=10,~S_{max}=50)$ and \textbf{B} for $(S_{min}=20,~S_{max}=100)$.}\label{fC}
\end{figure}

In FIG.~\ref{fC} the performance of the algorithm achieved by this improvement is illustrated. The network parameters are the same as FIG.~\ref{fb}. Now the method has better performance and the negative effect of heterogeneity of the community sizes is appreciably decreased. This is because even if some clusters are close to each other in the origin of projection space, the nodes in this region have less interactions with the rest of the network, especially with nodes other than their own community, hence they are more distant from nodes of  other clusters. Accordingly, these clusters are not mixed any more with the new definition of similarity. In addition, our method still works better for lager networks of $5000$ nodes. In the inset of FIG.~\ref{fC} the performance of the algorithm by using the Laplacian matrix instead of clumpiness matrix is illustrated. In small mixing parameters both matrices work well, however in large mixing parameters clumpiness matrix has a better performance.

The networks constructed for FIG.~\ref{fC} have the same parameters as the LFR networks used in the second figure of Ref.~\cite{CDA} to compare the performance of several methods. Hence, a comparison between our method and some other methods with high performances is possible. In comparison with the algorithm of Donetti and Mu\~{n}oz (DM) \cite{Donetti}, our algorithm has an appreciably better performance for $\mu\geq 0.6$. For $\mu<0.6$ our method and DM has nearly the same performance except for the $\lbrace N=5000, S\rbrace$ networks in which DM shows a considerable decreased performance. Compared to the inset of FIG.~\ref{fC}, the DM algorithm should work better in this region. This difference can be due to the error in finding the number of communities and the differences in employing hierarchical clustering algorithm. Additionally it is likely that a local maxima of modularity is achieved in the realizations of the DM method, as in the case of $\lbrace N=5000, S\rbrace$ networks the number of communities is much more, one should consider a larger maximum number of communities ($D$) in the DM method to obtain the global maxima of modularity.

In Ref.~\cite{CDA} the infomap method \cite{infomap} was shown to have a great performance with $I(A,B)=1$ up to $\mu=0.6$ for all networks, however it yields $I(A,B)\simeq 0$ for $\lbrace N=1000, B,~\mu\geq 0.65\rbrace$, $\lbrace N=1000, S,~\mu\geq 0.75\rbrace$, and $\lbrace N=5000, B,~\mu\simeq 0.8\rbrace$, and $I(A,B)<0.4$ for $\lbrace N=5000, S,~\mu\simeq 0.8\rbrace$. In comparison, our method yields values of $I(A,B)$ close to $1$ (the least $\langle I(A,B)\rangle\simeq 0.95$) up to $\mu=0.5$ and in large values of mixing parameter when infomap yields very small values of $I(A,B)$ our method has an appreciably better performance.

\subsubsection{Real-world examples}
In this part the method is used to illustrate the communities in several real networks which have $3$ or more communities.

\subsubsection*{\small Protein-protein interaction network}

\begin{figure} \centering
\subfloat[]
{\includegraphics[height=55mm]{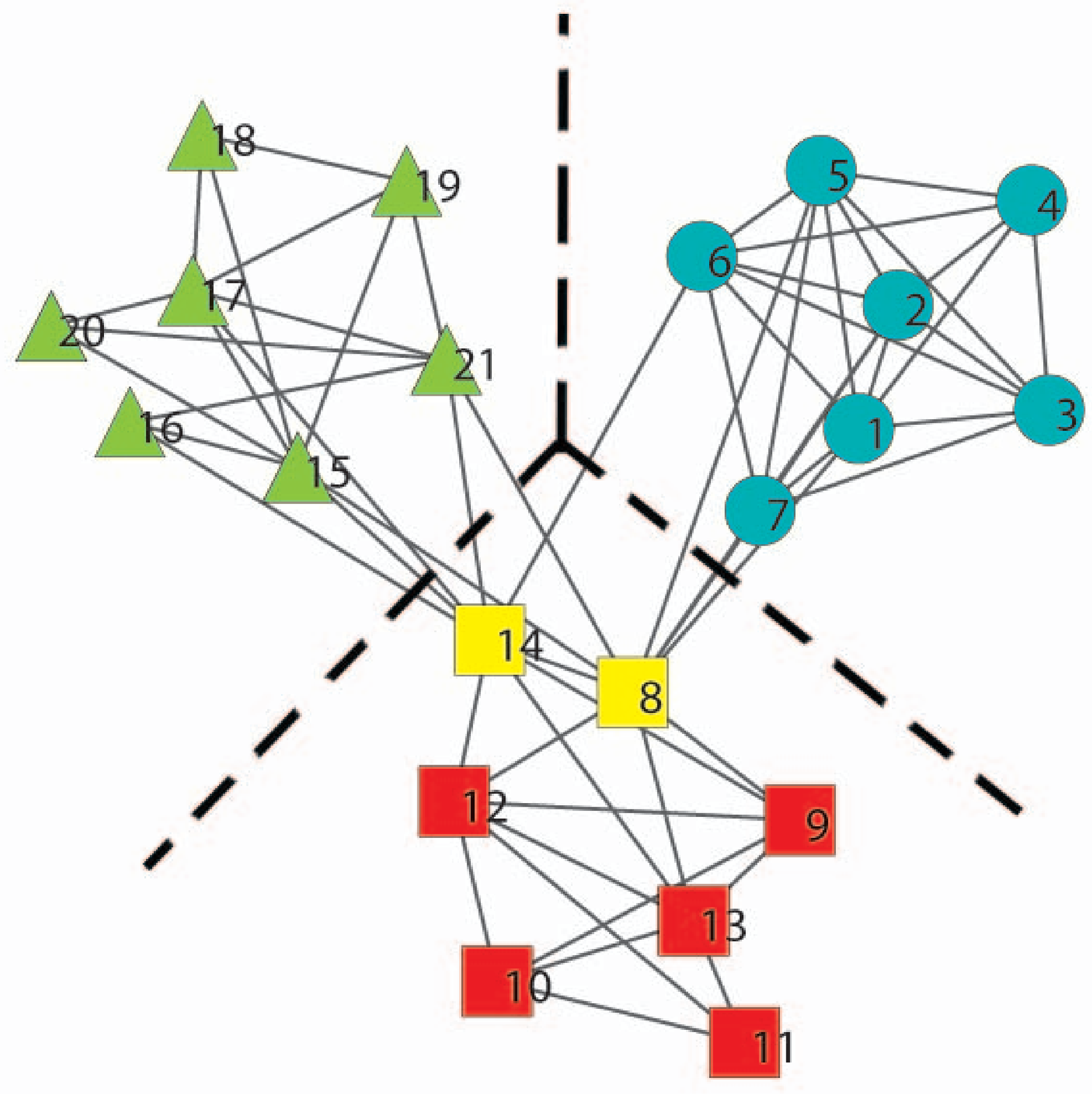}\label{fpa}}
\qquad
\subfloat []
{\includegraphics[height=55mm]{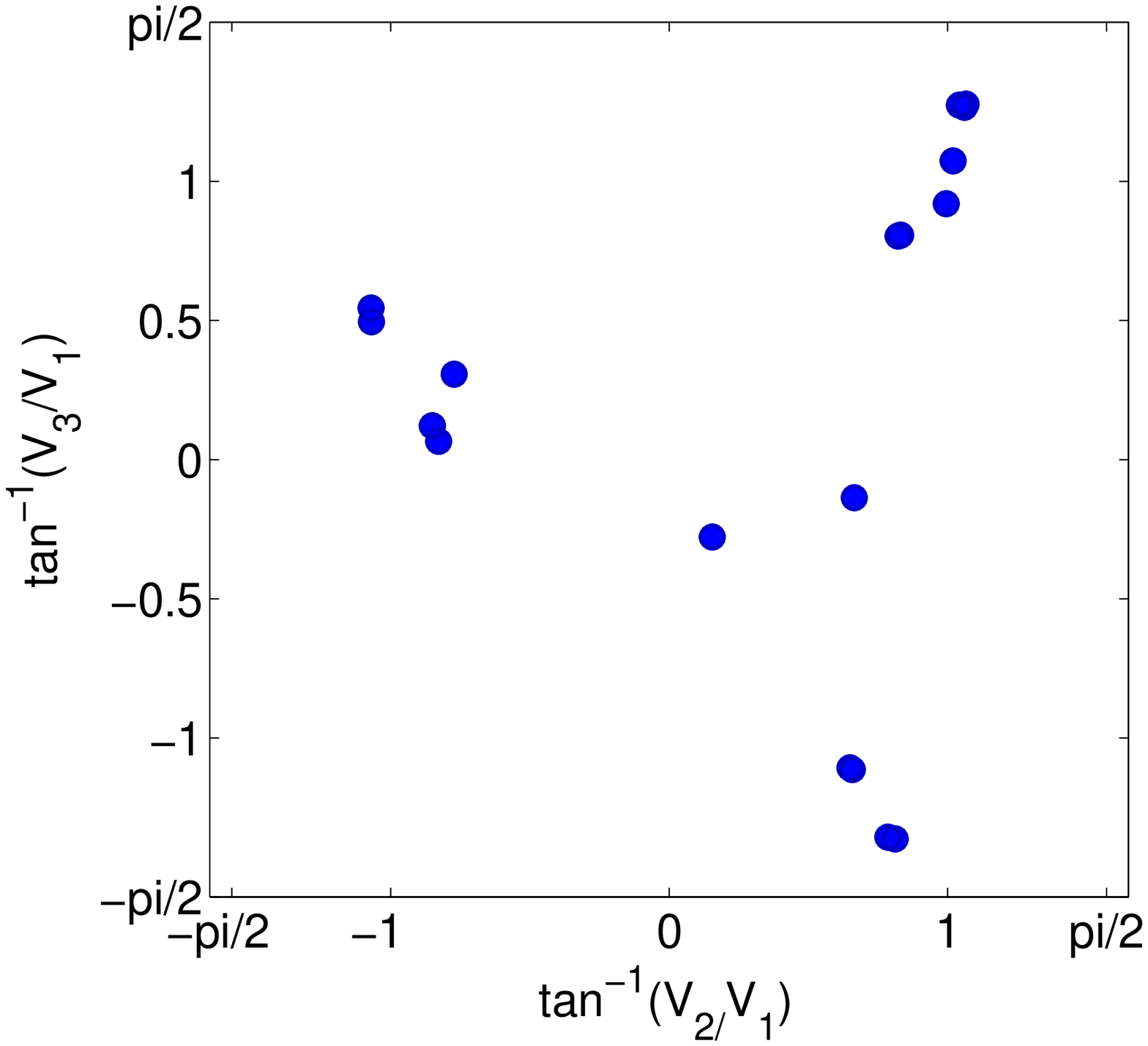}\label{fpb}}
\caption{(a) A graph constructed from interactions between proteins. This network is a sub-graph of S.cerevisiae protein-protein interaction network \cite{DIP} and shows the three communities assigned to the protein Zds1 (node number $8$ showed by a yellow square). (b) corresponding 2-D projection space. The dashed lines in (a) indicate the three detected communities which are in accordance with real data.}\label{fp}
\end{figure}

\normalsize
As our first example, we considered communities in a graph constructed from interactions between proteins (FIG.~\ref{fpa}) \cite{protein}. This biological network is comprised of three communities and members of each community either have a certain function or correspond to a protein complex \cite{protein}. The $\mathbf{\Gamma}$ projection space of this network is $2$-dimensional (FIG.~\ref{fpb}). The accumulation of points in three groups can be observed in FIG.~\ref{fpb}, which is in accordance with real communities. 

\subsubsection*{\small Citation network}

\begin{figure} \centering
\scalebox{.4}{\includegraphics{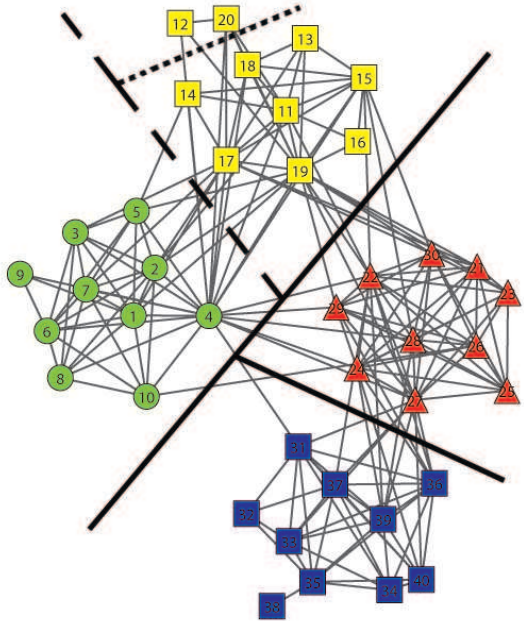}}
\caption{\label{fc} Community structure of a citation network. The solid and dashed lines indicate the partitions resulted from our algorithm given the number of communities to be $3$ or $4$ respectively. The solid line also shows the partitions found by infomap method. The additional dashed and dotted lines indicate the result of infomap method on the weighted version of the network. }
\end{figure}

\normalsize
The network of journal citation reports~\cite{citation,infomod} is an example of the case which we can have a good guess at the number of communities. This informational network is constructed of $40$ journals in four fields of physics (circles), chemistry (yellow squares), biology (triangles), and ecology (dark blue squares). In FIG.~\ref{fc} we showed the partitioning resulted from the infomap method~\cite{infomap} and our method by guessing the number of communities to be $3$ or $4$. The infomap code separates the network into $3$ communities, however by taking into account the weights of the links it detects $4$ communities correctly, apart from an excess group of $2$ nodes. Our algorithm detects the communities precisely according to the given number of them.

\subsubsection*{\small American collage football network}

\begin{figure} \centering
\subfloat[]
{\includegraphics[height=75mm]{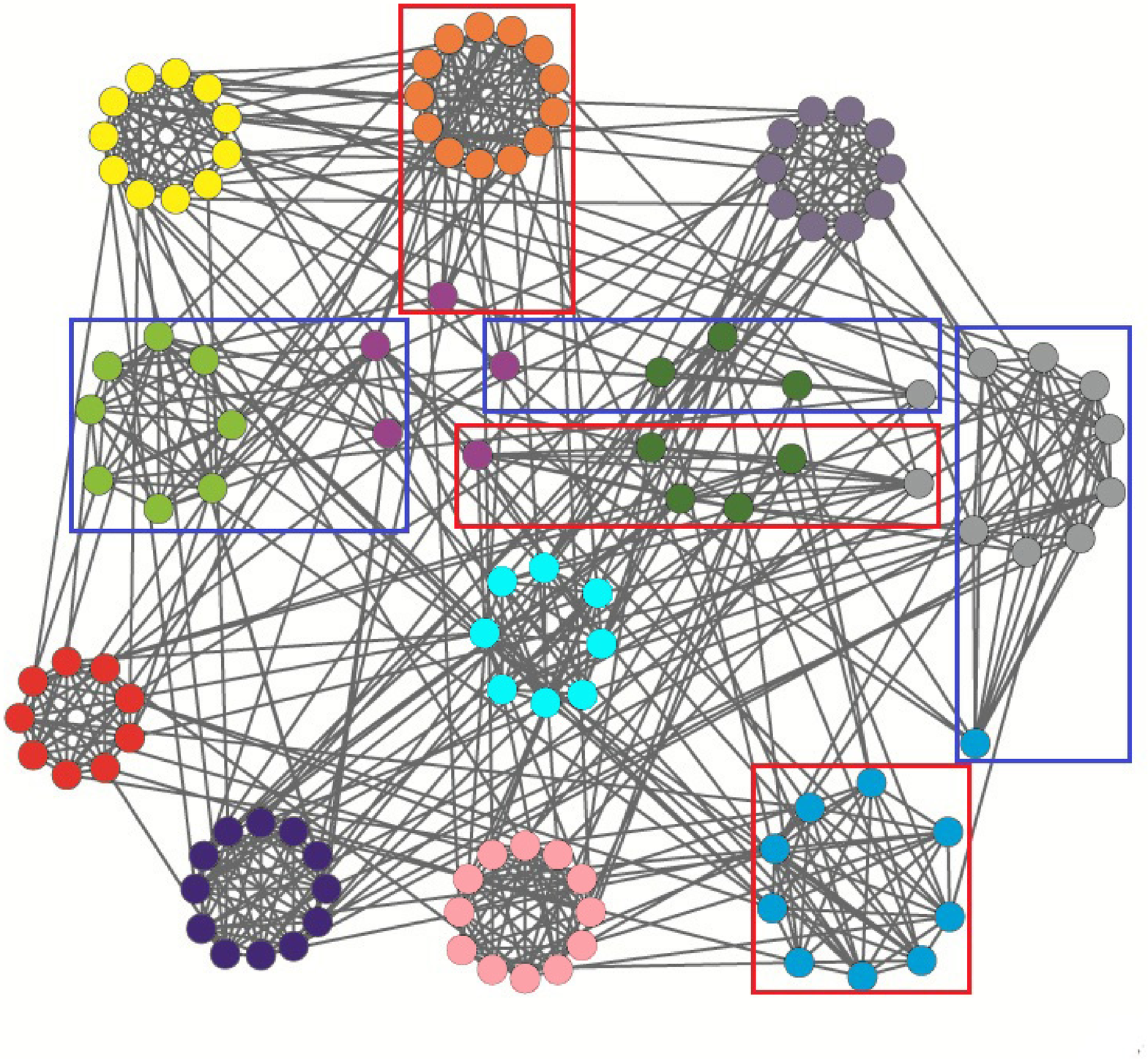}\label{ffa}}
\qquad
\subfloat []
{\includegraphics[height=75mm]{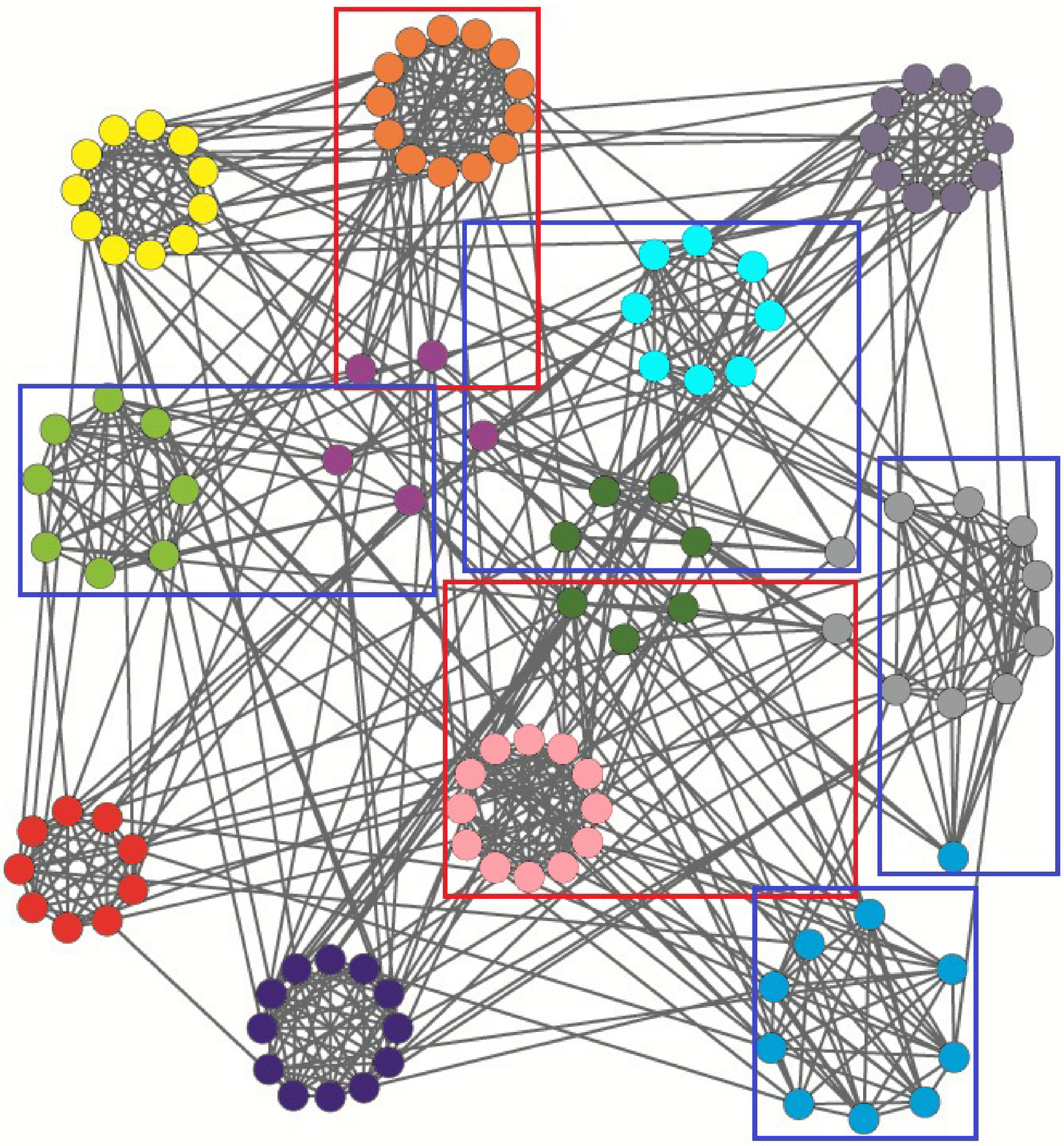}\label{ffb}}
\caption{The American collage football network \cite{football}. The teams of each conference are shown with distinct colors and gathered in circles. The square boxes show the partitions which were different from conferences. Other communities are identical with conferences. (a) Communities detected by considering 12 communities for the network. (b) Partitioning the network into 10 communities yields the maximum modularity of $Q=0.6046$.}\label{ff}
\end{figure}

\normalsize
Another example is a social network of American collage football teams (FIG.~\ref{ff}) \cite{football}. In this network the links show the games between 115 teams in year 2000. The teams were grouped in 12 conferences and most of the games were held between the teams of the same conferences. FIG.~\ref{ffa} shows the result of our algorithm when the network is assumed to have 12 communities; the modularity \cite{Newman} for this partitioning is $Q=0.6005$ and by assuming each conference as a real community the algorithm gives a normalized mutual information equal to $I(A,B)=0.9242$. The infomap method identifies 12 communities in the network and gives the same value for $I(A,B)$. However the maximum modularity is obtained for a partitioning with 10 communities ($Q=0.6046$) that yields a normalized mutual information equal to $I(A,B)=0.9522$.

\subsubsection*{\small Word association network}

\normalsize
Our final example is a graph derived from the word-association network of University of South Florida Free Association Norms \cite{WN}. The links between words in this network are constructed using an experiment. In the experiment, \textit{cue} words were given to 6000 participants and they were asked to introduce the first word that came to their mind (\textit{target} word) after seeing each cue word. There is a link between two words if they have a cue-target or target-cue relation.

\begin{figure} \centering
{\includegraphics[height=80mm]{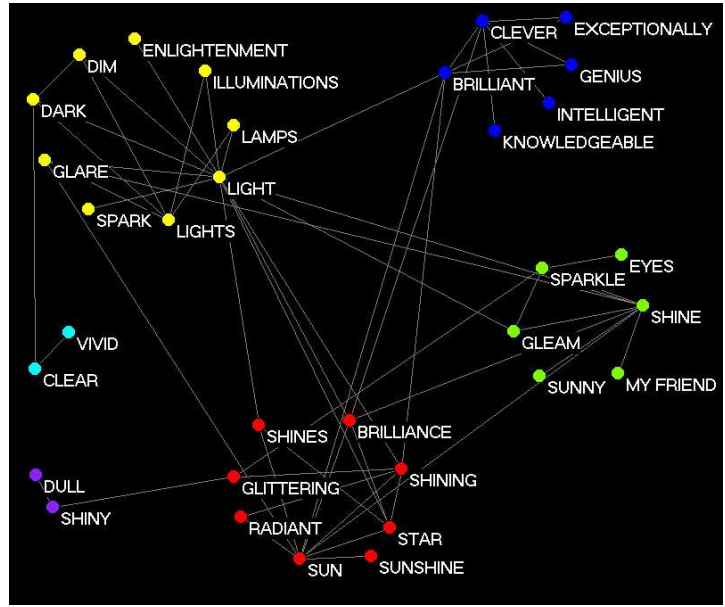}}
\caption{A word graph constructed from the relations between neighbors of the word \textit{bright} in a free association network. The nodes of each community, detected by our method, are plotted close together and have the same color. There are close relations between words of the graph, however words of each community are practically more similar.}\label{fw}
\end{figure}

The nodes in our graph (FIG.~\ref{fw}) are the neighbors of the word \textit{bright} and there is a link between two words if one of them is mentioned at least 3 times as the target word. There are close relation between theses words (i.e. semantic, syntactic, etc), however the graph can be partitioned to obtain groups of words which are practically more similar and have closer function. FIG.~\ref{fw} shows a good partitioning of the graph into 6 communities with modularity $Q=0.4479$ resulted from our algorithm.

\section{Concluding remarks\label{s4}}
In this paper we introduced a method for detecting communities of networks based on spectral properties of clumpiness matrix and the projection space corresponding to eigenvectors of this matrix. Our analysis shows that the method gives accurate results for many computer-generated and real-world networks. In benchmark networks, the method maintain a good performance even in large mixing parameters. The accuracy of the method was checked by normalized mutual information. The method finds communities in just one step not in a hierarchical manner. We also discussed qualitatively the effectiveness of the method by looking at the clumpiness matrix as the Hamiltonian of the system. The effect of size heterogeneity is also discussed and tested on some benchmarks. It is also observed that the algorithm gives better results for larger networks.

It is observed that the method can be appreciably improved by employing a minor change in the hierarchical clustering method. The improvement of algorithm in other aspects is also possible. For example one can try to build a graph matrix whose eigenvectors can better highlight the communities. This can be done by substituting another centrality measure (e.g. betweenness, closeness, random walk centrality \cite{network}) instead of nodes' degree, which is more related to the role of nodes in communities. Additionally other suitable distance measures can be used (e.g. one based on random walks introduced in \cite{Latapy}). One can also change the exponent of the $d_{i,j}$ in the elements of the clumpiness matrix and assign an arbitrary value $\alpha$ to it. By the current distance definition our observations, show that the effect of varying $\alpha$ between $0.5$ and $3.5$ on $I(A,B)$ is not greater than $\Delta(I(A,B))=0.02$.

It is worth mentioning that the extension of our method to the case of weighted networks is also possible. To this end one should replace the degree of a node with the sum of its links' weights. The shortest path length between two nodes can be defined in various ways depending on the structure of the network and meaning of weights. This is the subject of further investigation.

\section*{Acknowledgement}
We would like to thank M Zarei for useful discussions and remarks and introducing some references. We would also like to thank M Amini for his help with numerical simulations.

\end{document}